\documentclass[prd,preprint,a4paper,nofootinbib]{revtex4}
\usepackage{graphicx}
\usepackage{amsmath}
\usepackage{amsfonts}
\usepackage{amssymb}
\usepackage{bm}

\def\ba{\begin{eqnarray}}
\def\ea{\end{eqnarray}}
\def\bq{\begin{equation}}
\def\eq{\end{equation}}

\def\sla#1{\ifmmode%
\setbox0=\hbox{$#1$}%
\setbox1=\hbox to\wd0{\hss$/$\hss}\else%
\setbox0=\hbox{#1}%
\setbox1=\hbox to\wd0{\hss/\hss}\fi%
#1\hskip-\wd0\box1 }

\newcommand{\ttb}{$t\bar t$ }

\begin{document}
\thispagestyle{empty}

\preprint{
\hbox to \hsize{
\hfill\vtop{\hbox{Edinburgh 2002/24}
            \hbox{December 2002}} }
}

\title{\vspace*{.45in}
Top Pair Production Beyond Double-Pole Approximation:\\$\bm{p}\bm{p}$,\ $
\bm{p}\bm{{\bar{p}}} \bm{\to}$ 6 Fermions and 0, 1 or 2 Additional Partons }

\author{\vspace*{3mm}N.~Kauer}
\affiliation{\vspace*{3mm}
School of Physics, University of Edinburgh, Edinburgh EH9 3JZ, UK
\vspace*{.45in}}

\renewcommand{\abstractname}{{\normalsize \bf Abstract}}
\begin{abstract}
Hadron collider cross sections for $t\bar t$ production and di-lepton, single-lepton
and all-jet decays with up to 2 additional jets are calculated using complete LO
matrix elements with 6-, 7- and 8-particle final states.
The fixed-width, complex-mass and overall-factor schemes (FWS, CMS \& OFS) are employed
and the quality of narrow-width and double-pole approximations (NWA \& DPA) is
investigated for inclusive production and suppressed backgrounds to new particle
searches.  NWA and DPA cross sections differ by 1\% or less.
The inclusion of sub- and non-resonant amplitudes effects a cross section increase of
5--8\% at $pp$ supercolliders, but only minor changes at the Tevatron.
On-shell $t\bar t/Wtb$ backgrounds for the $H\to WW$ decay in weak boson fusion, the hadronic $\tau$ decay of a heavy $H^\pm$ and the $\phi\to hh\to\tau\tau b\bar{b}$ radion
decay at the LHC are updated, with corrections ranging from 3\% to 30\%.
FWS and CMS cross sections are uniformly consistent, but OFS cross sections
are up to 6\% smaller for some backgrounds.
\end{abstract}

\maketitle



%



\section{Introduction\label{intro}}

Since the discovery of the top quark in 1995 \cite{CDFD0}, the capabilities
of ongoing and forthcoming collider experiments have improved significantly.
Consequently, \ttb production will be abundant and studied intensely
as a signal at Fermilab's Tevatron collider and even more so at CERN's Large
Hadron Collider (LHC).  With a decay width $\Gamma_t$ of about 1.5 GeV the
top quark decays too rapidly to be observed directly, and is instead 
identified through characteristic detector signatures with isolated leptons
and jets.  These signatures would also be observed in the production of various
hypothetical particles, so that top production constitutes an important
background for many new particle searches.  In light of the changed role that
top production will play in the near future, the quality of corresponding theoretical
predictions needs to be reviewed.

As is well known, a general, systematic and ``natural'' treatment of unstable
particles in perturbative field theory is not straightforward.\footnote{Even limited,
appealing schemes like the fermion-loop scheme
become rather involved for all but the simplest
applications \protect\cite{Baur:1995aa,fermionLoopScheme}.}
Signal cross sections that are dominated by the
production and decay of unstable particles with $\Gamma/m\ll 1$ can be calculated
with good accuracy in narrow-width approximation (NWA).
This and similar approximations, like the leading-pole
approximation, focus on contributions on or close to resonance and thus greatly
simplify calculations, since the production and decay of unstable particles
(largely) factorises.  They have been widely employed to predict inclusive and
exclusive cross sections.  Their use to
determine background rates for experiments with restrictive selection cuts
that eliminate resonant contributions and emphasise peripheral phase space regions
can be problematic.  In such cases, users of general-purpose event generators like
PYTHIA \cite{pythia} and HERWIG \cite{herwig}
have applied a suggestive procedure that combines results in NWA at the cross
section level.  For \ttb backgrounds,
for example, \ttb and $Wtb$ results are added to account for double- and
single-resonant contributions.
Since \ttb calculations in NWA implicitly contain
sub- and non-resonant contributions that have been integrated out,
this procedure can lead to significant double-counting \cite{ttj}.
It also neglects interference effects:  top pair production
and associated top ($Wtb$) production are specialisations of one and the same
process, since initial and final states of both ``processes'' are identical.
Furthermore, calculations in NWA often do not include full spin correlations.

Examples of new particle searches at present and
future hadron colliders with substantial \ttb and $Wtb$ backgrounds include
$H\to W^+W^-$~\cite{DittDrein,CMS,ATLAS_TDR_2,Trefzger_Higgs,KPRZ,RZ_WW} 
and $H\to\tau^+\tau^-$~\cite{ATLAS_TDR_2,RZH_tau,PRZ_TauTau} decays,
leptonic signals for cascade decays of supersymmetric particles~\cite{SUSY},
the Randall-Sundrum radion decays
$\phi\to W^+W^-\to \ell^+\ell^-\nu\bar{\nu}$ \cite{radionWW}
and $\phi\to hh\to b\bar{b}\tau^+\tau^-$ \cite{radiontautau},
and searches for $H^-\to \tau_L^-\nu$ in models with a singlet
neutrino in large extra dimensions \cite{chargedHiggsExtraDim}.
Reliable phenomenological studies of these searches require tools that allow
accurate calculations of top pair production and decay in resonant as well
as non-resonant phase space regions and that are not susceptible to the shortcomings
mentioned above.  The calculations should therefore employ complete
matrix elements, i.e.~the sum of all leading-order (LO) amplitudes.
These matrix elements -- in fact all resonant fixed-order amplitudes -- exhibit
unphysical singularities, and a finite-width scheme has to be applied to reflect
that in field theory propagators of unstable particles acquire complex poles
when self-energies are resummed to all orders.  The set of higher-order
contributions that has to be included to adequately model finite-width effects is not
uniquely determined.  Moreover, variations of higher order in
$\Gamma/m$, as well as exclusion of problematic phase space regions, e.g. thresholds,
are permissible. Consequently, a variety of competing schemes exists
\cite{simpleFiniteWidthSchemes,overallFactorScheme,complexMassScheme,Baur:1995aa,fermionLoopScheme}.

The purpose of this paper is to compare leading-order
\ttb cross sections calculated
in narrow-width or double-pole approximation (DPA)
to cross sections that take into account all sub- and non-resonant amplitude
contributions, and to investigate the consistency of several
practical finite-width schemes.
The program we developed for this purpose is described in Section \ref{program},
with particular emphasis of finite-width schemes and their implementation.
In Section \ref{inclusive} results for inclusive top pair production are
presented, followed by results for important top backgrounds to new particle
searches in and beyond the Standard Model (SM) in Sections \ref{SMSearch}
and \ref{BeyondSM}, respectively.
In Section \ref{summary}, we conclude with a summary and outlook.

\section{\label{program}Program Description}

We introduce a LO program for \ttb production at hadron colliders
with up to two additional jets that is not specialised to
resonant phase space regions and hence has to include 
complete tree-level matrix elements for the contributing
$2\to 6$, $2\to 7$ and $2\to 8$ subprocesses.
If the $W^+W^-$ decay products are abbreviated as ${\cal W}$, 
\ttb production includes the subprocesses
\bq
gg \to b\bar{b}\; {\cal W}\,,\ \
q \bar{q}  \to b\bar{b}\; {\cal W}\,,
\eq
\ttb + 1 jet production includes the subprocesses
\bq
gg \to b\bar{b}\; {\cal W}\; g\,,\ \
q \bar{q}  \to b\bar{b}\; {\cal W}\; g\,,\ \
q g  \to b\bar{b}\; {\cal W}\; q\,,\ \
\bar{q} g \to b\bar{b}\; {\cal W}\; \bar{q}\,,
\eq
and \ttb + 2 jet production includes the subprocesses
\ba
& gg \to b\bar{b}\; {\cal W}\; gg\,,\ \ 
q \bar{q}  \to b\bar{b}\; {\cal W}\; gg\,,\ \
q g  \to b\bar{b}\; {\cal W}\; qg\,,\ \
\bar{q} g \to b\bar{b}\; {\cal W}\; \bar{q}g\,, & \nonumber\\
& gg  \to b\bar{b}\; {\cal W}\; q \bar{q}\,,\ \
q\bar{q}  \to b\bar{b}\; {\cal W}\; q\bar{q}\,,\ \
qq  \to b\bar{b}\; {\cal W}\; qq\,,\ \
\bar{q}\bar{q}  \to b\bar{b}\; {\cal W}\; \bar{q}\bar{q}\,. &
\ea
The program contains subprocess matrix elements for the
di-lepton, single-lepton and all-jet decay modes, or more
specifically for the following $W^+W^-$ decay final states:
\ba
{\cal W}_\text{di-lepton} & = & \ell^+\nu\;\ell^-\nu\,, \\
{\cal W}_\text{single-lepton} & = & \ell^+\nu\; q_d\bar{q}_u\,, \\
{\cal W}_\text{all-jet} & = & q_u\bar{q}_d\;q_d\bar{q}_u\,.
\ea
For the di-lepton and all-jet decay modes, the program allows to
calculate different-flavor samples,
e.g.~with ${\cal W} = e^+\nu_e\;\mu^-\bar{\nu}_\mu$,
as well as same-flavor samples,
e.g.~with ${\cal W} = e^+\nu_e\;e^-\bar{\nu}_e$.
Additional amplitudes with
($\gamma,Z\to\ell^+\ell^-) \times (Z\to\nu_\ell\bar{\nu}_\ell)$ and
$(g,\gamma,Z\to q_u\bar{q}_u)\times (g,\gamma,Z\to q_d\bar{q}_d)$
fragments contribute in the di-lepton and all-jet decay modes, respectively.
Moreover, a finite-width scheme has to be chosen when formulating
complete LO matrix elements with unstable particles to avoid
unphysical singularities in resonant phase space regions that
can be removed by including contributions to all orders in
perturbation theory.\footnote{The Dyson resummation of top quark
self-energy contributions is described in Ref.~\cite{ttj}.}
Since no known scheme is satisfactory in every respect, a cross
section by cross section comparison of several schemes with
complementary properties is suggestive, but requires more than
one version of each subprocess matrix element defined above.

Evidently, the creation of all
required matrix elements is a considerable task and calls for
automation.  While the program is generally written in
C++ to permit greater code locality and expressiveness, we
prefer faster Fortran code for the matrix element evaluation,
since its speed determines the program runtime after initial
adaptation.  Furthermore, to minimize the matrix element code,
the program should use helicity amplitudes in unitary gauge that
neglect CKM mixing.  MADGRAPH/HELAS \cite{MADGRAPH,HELAS} is a
matrix element generation system that matches our requirements
and has recently been extended to processes with 8-10 external particles.
Its output is used as starting point for the matrix element code
in our program.

MADGRAPH/HELAS matrix elements use the fixed-width scheme (FWS).\footnote{
We use HELAS-3, which implements the fixed-width scheme.
Note that the widely-used version 2 of HELAS implements step-width Breit-Wigner
propagators, i.e. $1/(p^2-m^2+im\Gamma\:\theta(p^2))$.
No notable deviations occur in general, since $|p^2-m^2|\gg m\Gamma$ if 
$p^2 < 0$ and $\Gamma/m \ll 1$.
}
In the FWS, all propagators of unstable particles are modified according to the
following prescription:\bq
\frac{1}{p^2-m^2} \to \frac{1}{p^2-m^2+im\Gamma}\,.
\eq
This substitution is easy to implement, but the resulting matrix
elements with Breit-Wigner propagators are not gauge-invariant.  As discussed in
Refs.~\cite{Baur:1995aa,fermionLoopScheme,ttj}, calculations that employ
gauge-variant amplitudes and receive sizable contributions from
sensitive phase space regions can yield highly erroneous results.
To remedy this deficiency, various approaches have been suggested
in the literature that yield manifestly gauge-invariant matrix elements.
The theoretically most appealing approach is arguably the fermion-loop
scheme \cite{Baur:1995aa,fermionLoopScheme}.  We do not consider it further
here, since it is not applicable to processes with unstable
particles that decay into bosons, including \ttb production.
Even if it were applicable in the case at hand, it would require
as prerequisite an analytic calculation of effective vertices
that has not been automatised yet.  Its implementation is therefore
not straightforward for complex multi-particle processes with
several types of unstable particles.  For the studies in Section \ref{results}
we therefore implement two practical finite-width schemes that
allow automatic matrix element generation for arbitrary processes
and guarantee electroweak and $SU(3)$ gauge-invariant results:
the complex-mass scheme (CMS) \cite{complexMassScheme}
and the overall-factor scheme (OFS) \cite{overallFactorScheme}.

The CMS introduces Breit-Wigner
propagators in a gauge-invariant manner by replacing the
masses of all unstable particles with a complex value as follows:
\bq 
m \to \sqrt{m^2 - i m \Gamma}\,.
\eq
This substitution is performed unconditionally and
yields, for example, for the top propagator
a different expression than the FWS:
$i(\sla{p} + \sqrt{m_t^2 - i m_t \Gamma_t})/(p^2-m_t^2+im_t\Gamma_t)$.
$\sin^2\theta_W$ and dependent quantities also acquire complex
values in this scheme, since $\cos\theta_W = m_W/m_Z$.
The CMS matrix elements in our program use HELAS-CMS, a modified
version of the HELAS library that we created by converting masses
and widths from real to complex variables.\footnote{Complex widths are
  introduced since MADGRAPH output uses a real constant ZERO for
  both, vanishing masses and widths, as argument for HELAS calls.
  In CMS matrix element code we define ZERO as complex parameter
  and then also have to declare all widths as complex variables
  to be compatible.  Note that all width variables are set to zero in CMS
  matrix elements, since the widths are contained in the mass variables.}

The OFS conserves gauge-invariance while introducing Breit-Wigner behaviour
by multiplying the complete LO matrix element (with singular
propagators for unstable particles) with overall factors:
\bq
{\cal M}_{compl.} \times  \frac{p^2 - m^2}{p^2 - m^2 + i m    \Gamma}\
    = {\cal M}_{res.,\text{BW-prop.}}\,+\, {\cal M}_{non-res.} \times
    \frac{p^2 - m^2}{p^2 - m^2 + i m \Gamma}\,.
\label{OFSEquation}
\eq
For each unstable particle type, one factor is applied for every
time-like momentum combination that occurs in propagators of that
type.  The propagators absorb
the corresponding factor and transform into Breit-Wigner propagators:
\bq
\frac{1}{p^2 - m^2}\,\times\, \frac{p^2 - m^2}{p^2 - m^2 + i m \Gamma}\ 
    \,\to \ \,\frac{1}{p^2 - m^2 + i m \Gamma}\,.
\eq
Amplitudes that are non-resonant with respect to a particular momentum
combination do not absorb the corresponding factor, as indicated
in Eq.~(\ref{OFSEquation}).
To facilitate the automatic construction of OFS matrix elements
a scripting-language program was written that scans MADGRAPH output,
and analyses the structure of all contributing amplitudes.  Potentially
resonant propagators, where one side is only connected to final state particles,
are identified and the required overall factors deduced.  The script then
constructs the overall factor product for each amplitude, and outputs Fortran code
that calculates the OFS matrix element.  To optimise the code, combinations of
overall factors that occur multiple times are evaluated once and the results
are reused.

\label{implementation}
The comparisons presented in Section \ref{results} also require the calculation
of cross sections in double-pole and narrow-width approximation.
For that purpose, a second program -- in nature similar to the one used to generate
OFS matrix elements -- eliminates all amplitudes that do not contain potentially
resonant $t$ as well as $\bar t$ propagators, thus extracting all double-resonant
amplitudes
with respect to top decay.  The generated Fortran code employs the fixed-width scheme
and is used in our DPA calculations.
The DPA matrix elements are also used in our NWA calculations.  To preserve all spin
correlations, we choose to implement the NWA directly by calculating with off-shell
intermediate top quarks in the $\Gamma_t\to 0$ limit.
To that end, the top width is scaled down to
$\Gamma_{t,\text{eff}} = \varepsilon \Gamma_t$, and $|{\cal M}|^2$ 
is multiplied by $\varepsilon^2$ to restore the proper normalisation of the total amplitude.  For one resonant propagator one has
$|{\cal M}_{\text{eff}}|^2 = 1/\varepsilon\times|{\cal M}|^2$.
A setting of $\varepsilon = 1/1000$ is used in the program and yields excellent
agreement with NWA implementations with on-shell intermediate top states.
In DPA or NWA mode, the program uses a Breit-Wigner mapping for each
resonant top propagator that covers a limited range of invariant top quark masses.
Neglected contributions from outside this range introduce a non-statistical error.
For the background calculations in Sections
\ref{SMSearch} and \ref{BeyondSM}, off-shell top masses were generated in 
$m_t \pm 65 \Gamma_t$ limiting neglected contributions to approximately 1\%
(see Eq.~(19) in Ref.~\cite{ttj}).  For the inclusive calculations in
Section \ref{inclusive}, we increased the range factor to 6500,
reducing this error contribution to 0.01\%.
A comparison of results given in Table
\ref{DiLeptonInclusiveComparisons} below with results 
in Table II in Ref.~\cite{ttj} confirms that a range factor of 65
is not sufficient when a total error of less than 1\% is desired.

When cross sections for \ttb production with additional jets
are calculated with complete matrix elements,
one finds that computational complexity increases by a factor of more than
10 for each additional final state particle beyond the \ttb level.
Resulting program runtimes quickly exceed what would be considered acceptable for
phenomenological studies.  To obtain the $t\bar{t}jj$ results presented in
Section \ref{SMSearch}, it was therefore necessary to use
state-of-the-art integration techniques and to develop a method to
distribute the Monte Carlo sampling over a larger number of processors.
The result is OmniComp, a Monte Carlo integration framework based on
the adaptive multi-channel techniques introduced in
Refs.~\cite{VEGAS,KleissMultiChannel,OhlVEGAS} that allows to conveniently
distribute the calculation over many processors in one or more computer
clusters.\footnote{We successfully ran programs on up to 16 processors.}
OmniComp further accelerates the computation of hadron collider cross sections
through adaptive Monte Carlo summation of subprocess $\bm{\otimes}$ helicity
combination channels.  The mapping of sub- and non-resonant phase
space regions follows the approach laid out in Ref.~\cite{KauerThesis}.
OmniComp and the phase space mapping library are described in more detail in
Ref.~\cite{TechnicalWriteUp}.  A number of tests were applied to verify the
correctness of the program.  First, the Lorentz-invariance of
the MADGRAPH-generated FWS matrix elements was tested.\footnote{
The Lorentz-invariance of one and the equivalence of two matrix element routines
was tested as described in Ref.~\protect\cite{ttj}.}
CMS matrix elements were tested by comparison with corresponding FWS matrix elements
after the complex masses and widths in HELAS-CMS had been set to their usual, real
values.  The automatic generation of OFS matrix elements was tested by
comparing with the manually created OFS matrix elements of Ref.~\cite{ttj}.
The DPA/NWA matrix elements were verified by comparing NWA cross sections
with results from programs with on-shell intermediate top quarks.
The phase space and PDF integration has been tested by comparing
with known cross sections for the LHC and Tevatron.
Moreover, the addition of hadron collider capabilities to the general purpose
packages O'Mega \& Whizard \cite{OMegaWhizard} and AMEGIC++ \cite{AMEGIC}
reached the final stage this year, and a comparison of top production cross
sections to cross-check our implementations is planned for the near future.


\section{Numerical Results\label{results}}

In this section, we use the program described above to study
the difference between cross sections with on-shell (NWA) and
off-shell (DPA) intermediate top quarks, to determine the size of
corrections when complete LO matrix elements are included, and to search
for deviations between results obtained with different finite-width schemes.\footnote{
Note that all results calculated with our program include full spin
correlations (see Section~\protect\ref{implementation}).}
We first investigate these issues for inclusive top pair production 
 and then turn to suppressed top backgrounds that are important 
for new particle searches in and beyond the Standard Model.

To cover the energy range of existing and future hadron
colliders, cross sections are calculated for the Tevatron ($p\bar p$, $\sqrt{s} = $ 2 GeV),
the LHC ($pp$, $\sqrt{s} = $ 14 TeV) and a Stage-1 VLHC ($pp$, $\sqrt{s} = $ 40 TeV).
Unless otherwise noted, all calculations use the following parameters: 
$m_Z =$ 91.187 GeV, $G_F = 1.16637\times 10^{-5}$ GeV$^{-2}$ and $\alpha(m_Z) = 1/128.92$,
which translates at tree-level to $\sin^2\theta_W$ = 0.23105 and $m_W$ = 79.9617 GeV,
as well as the masses $m_t$ = 175 GeV, $m_b$ = 4.4 GeV and $m_H$ = 115 GeV.  LO formulas for
the decay widths then yield $\Gamma_t = 1.56$ GeV, $\Gamma_W = 2.01$ GeV,
$\Gamma_Z = 2.42$ GeV and $\Gamma_H = 0.00323$ GeV.
CTEQ6L LO parton distribution functions are employed by default, with $\alpha_s(m_Z)$ = 0.118
and the NLO formula.  Factorisation and renormalisation scales are fixed at the top mass,
except for studies where cross sections with additional jets are taken into account
(e.g.~in Table \ref{LightHiggsDiscoveryOffshellTopBackgrounds}).  In this case, the
factorisation scale is chosen as $\mu_f = \min(m_T)$ of the top quarks and additional jets.
This factorisation scale definition avoids double-counting of contributions
that have already been integrated out in the PDFs.
The overall strong coupling constant factor is calculated as $(\alpha_s)^n = \Pi_{i=1}^n
\alpha_s(m_{T,i})$, again using the transverse masses of both top quarks and any additional
jets as input.  
ATLAS detector resolution and $b$ decay effects are modelled as described in
Ref.~\cite{PRZ_TauTau}, but tagging efficiencies are not taken
into account.
Monte Carlo integration errors are 0.1\% or less for inclusive cross sections
and 1\% or less for background cross sections.


\subsection{\label{inclusive}Inclusive Production}

\begin{table}
\caption{\label{DiLeptonInclusiveComparisons}
Cross sections in NWA and with complete matrix elements for
inclusive \ttb production and di-lepton decay ($b\bar{b}\: e^+ \nu_e\:
\mu^- \bar{\nu}_\mu$).
Effects are compared for colliders, PDF sets and
finite-width schemes.  
Tevatron corrections are of order $\Gamma_t/m_t = 0.009$, but LHC
corrections are larger.  PDF improvements decreased the LHC cross
section by 18\%, but the correction is robust.  FWS, CMS and OFS
yield consistent results.
All cross sections are given in fb.
Note that NWA cross sections include
full spin correlations (see Section~\protect\ref{implementation}).
}
\vspace*{.5cm}
\begin{tabular}{|l|c|c|c|c|c|}
 \hline
 Collider & $\sigma_\text{NWA}$ & $\sigma_\text{FWS}$ & $\frac{\sigma_\text{FWS}}{\sigma_\text{NWA}}$ & $\frac{\sigma_{gg,\text{FWS}}}{\sigma_{gg,\text{NWA}}}\ \ (\frac{\sigma_{gg,\text{FWS}}}{\sigma_\text{FWS}})$ & $\frac{\sigma_{q\bar{q},\text{FWS}}}{\sigma_{q\bar{q},\text{NWA}}}\ \ (\frac{\sigma_{q\bar{q},\text{FWS}}}{\sigma_\text{FWS}})$ \\
 \hline
 LHC      & $5.86\times 10^3$ & $6.19\times 10^3$ & 1.06 & 1.06 (88\%) & 1.00 (12\%) \\
 Tevatron & 66.5 & 66.5 & 1.00 & 1.10 (5\%) & 0.99 (95\%) \\
 \hline
\end{tabular}\\
\vspace{.5cm}
\begin{tabular}{|l|c|c|c|}
 \hline
 PDF set & $\sigma_\text{NWA}$ & $\sigma_\text{FWS}$ & $\sigma_\text{FWS}/\sigma_\text{NWA}$ \\
 \hline
 CTEQ6L & $5.86\times 10^3$ & $6.19\times 10^3$ & 1.06 \\
 CTEQ4L & $7.18\times 10^3$ & $7.58\times 10^3$ & 1.06 \\
 \hline
\end{tabular}\\
\vspace{.5cm}
\begin{tabular}{|l|c|c|}
 \hline
 Scheme & $\sigma$ & $\sigma/\sigma_\text{NWA}$ \\
 \hline
  NWA & $5.86\times 10^3$ & 1.00 \\
  FWS & $6.19\times 10^3$ & 1.06 \\
  CMS & $6.19\times 10^3$ & 1.06 \\
  OFS & $6.18\times 10^3$ & 1.06 \\
 \hline
\end{tabular}
\vspace*{.5cm}
\end{table}

\begin{table}
\caption{\label{InclusiveOffshellChangeComparison}
Change of cross sections with complete matrix
elements relative to cross sections in NWA 
for inclusive \ttb production and di-lepton, single-lepton
and all-jet decay at the 
Tevatron, LHC, and Stage-1 VLHC ($\sqrt{s} = $ 40 TeV).
At $pp$ supercolliders, cross sections increase uniformly by about 7\%
for all decay modes and collider energies.
Effects are similar for cross sections in DPA (shown
in parenthesis): replacing on-shell with off-shell intermediate
top quarks changes cross sections by 1\% or less.
The cuts $p_T > 15$ GeV, $|\eta| < 4.5$ and $\Delta R > 0.6$
are applied for channels with singular phase space
regions related to massless particles.
}
\vspace*{.5cm}
\begin{tabular}{|l|c|c|c|}
 \cline{2-4}
\multicolumn{1}{c|}{} & \multicolumn{3}{c|}{$\sigma_\text{CMS}/\sigma_\text{NWA}$ ($\sigma_\text{CMS}/\sigma_\text{DPA}$)} \\
 \cline{2-4}
\multicolumn{1}{c|}{}  & di-lepton & single-lepton & all-jet \\
 \hline
 Tevatron & 1.00 (1.01) & 1.01 (1.02) & 1.00 (1.01) \\
 LHC      & 1.06 (1.06) & 1.07 (1.07) & 1.07 (1.07) \\
 VLHC     & 1.06 (1.06) & 1.07 (1.06) & 1.08 (1.07) \\
 \hline
\end{tabular}\\
\vspace*{.5cm}
\end{table}

For Table \ref{DiLeptonInclusiveComparisons}, we choose inclusive \ttb
production and decay into the di-lepton final state $b\bar{b}\: e^+ \nu_e\:
\mu^- \bar{\nu}_\mu$, and compare cross sections in NWA to cross sections with
complete matrix elements.  First, the size of changes is compared for LHC 
and Tevatron collisions.  Generally one would expect finite-width effects to 
be of order $\Gamma_t/m_t = 0.009$, and the Tevatron correction is indeed less than
1\%.  The LHC cross sections, however, is enhanced by a significantly larger
factor of 1.06.  This effect is not caused by averaged v.~exact spin
correlations, since our NWA results include full spin correlations.
Furthermore, less than 1\% of the increase can be attributed to
double-resonant off-shell effects (as seen in Table
\ref{InclusiveOffshellChangeComparison}).  The increase is mainly caused by 
previously omitted sub-resonant contributions.  These contributions are also
included for the Tevatron.  That no sizable increase occurs there can be 
traced to the fact that the Tevatron cross section is dominated by quark 
scattering, while the LHC cross section is dominated by gluon scattering.
As shown in the two rightmost columns, the large increase is specific to the
gluon-initiated process.  In fact, the $q\bar q$ cross section is slightly
reduced when sub- and non-resonant amplitudes are included.
A suggestive kinematical interpretation that relates the large increase to off-shell
contributions at hard scattering energies below the on-shell top pair production
threshold that are amplified by steeply falling PDFs is therefore misleading.
A comparison of calculations with CTEQ4L and CTEQ6L shows that
recent PDF improvements decrease cross sections uniformly by 18\%,
thus having little effect on the $\sigma_\text{FWS}/\sigma_\text{NWA}$
enhancement factor.
For inclusive cross sections, differences between finite-width schemes are
expected to be of higher order in $\Gamma_t/m_t$.  The third comparison in Table
\ref{DiLeptonInclusiveComparisons} shows that FWS, CMS and OFS yield
results that agree when integration errors of 0.1\% are taken into account.

Cross section changes when progressing from NWA to DPA and finally to
complete LO matrix elements are displayed
in Table \ref{InclusiveOffshellChangeComparison}
for the di-lepton, single-lepton and all-jet channels of 
inclusive \ttb production at three hadron colliders:
Tevatron, LHC, and a Stage-1 VLHC with $\sqrt{s} = $ 40 TeV.
In all cases the ratios $\sigma_\text{CMS}/\sigma_\text{NWA}$
and $\sigma_\text{CMS}/\sigma_\text{DPA}$ are very similar:
Replacing on-shell with off-shell intermediate top quarks changes cross
sections by no more than 1\%.  Effects are generally small at the Tevatron.
At $pp$ supercolliders, on the other hand, cross sections increase uniformly by
about 7\% for all decay modes and collider energies.


\subsection{\label{SMSearch}Backgrounds to SM Higgs Searches}

\begin{figure}
\begin{minipage}[c]{.48\linewidth}
\flushright \includegraphics[width=6.cm, angle=90]{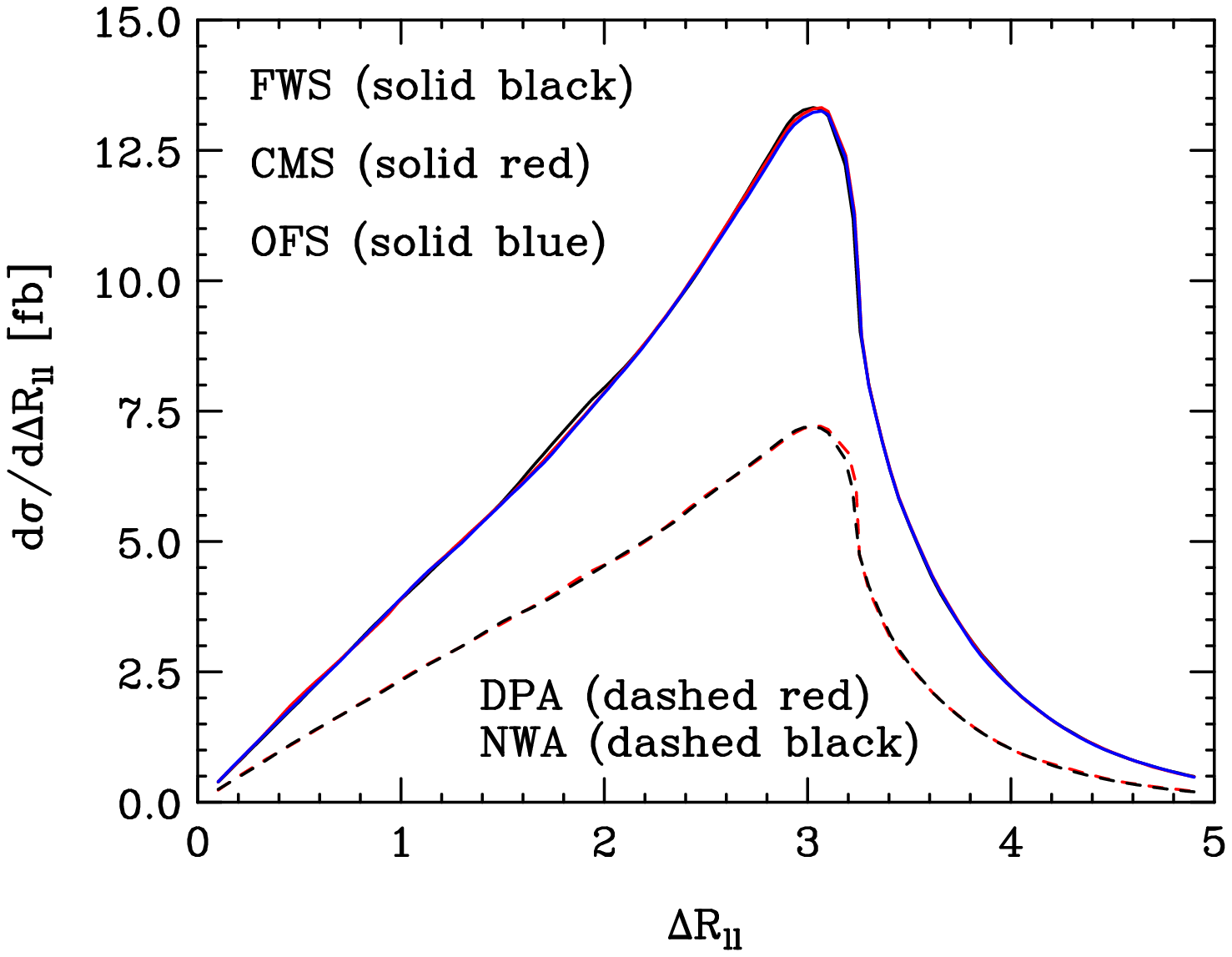}
\end{minipage} \hfill
\begin{minipage}[c]{.48\linewidth}
\flushleft \includegraphics[width=6.cm, angle=90]{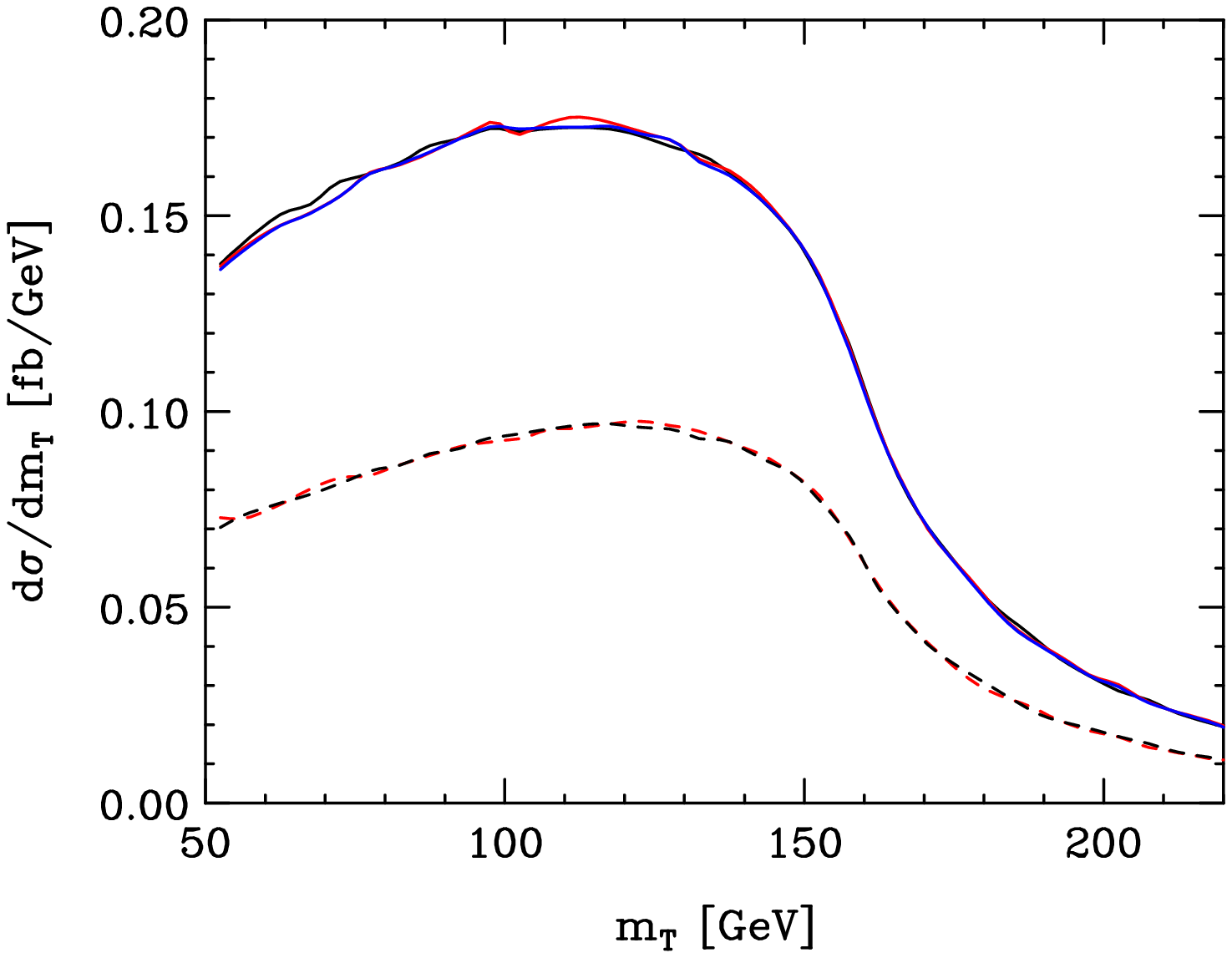} 
\end{minipage}
\caption{\label{JetVetoDistributions}
Charged lepton separation in $\eta$-$\phi$ space and transverse
mass distributions for suppressed \ttb production and di-lepton decay
($b\bar{b}\: e^+ \nu_e\:\mu^- \bar{\nu}_\mu$) at the LHC.
The applied central jet veto ($p_T > 15$ GeV and $|\eta| < 3.2$) reduces
the \ttb acceptance to $4\times10^{-3}$.  Differential cross
sections with complete matrix elements (solid lines) and
in NWA and DPA (dashed lines) are shown.  Sub- and
non-resonant amplitude contributions enhance the total cross
section by a factor 1.8.  Off-shell top effects (DPA v.~NWA) and deviations
between finite-width schemes (FWS, CMS, OFS) are negligible.
The transverse mass is defined as $m_T = \sqrt{2p_T^{\ell\ell}\sla E_T[1-\cos\Delta\theta(\ell\ell,\sla E_T)]}$.
}
\end{figure}

The unexpectedly large increase of inclusive cross sections at 
$pp$ supercolliders when sub- and non-resonant contributions are included
raises the question how large corresponding effects are for new particle
searches with significant \ttb backgrounds.  When optimised selection cuts 
are used to suppress top pair production, the dominant double-resonant \ttb
contributions are typically suppressed by factors of order $10^{-4}$ and
the importance of contributions from sub- and non-resonant phase space regions
can increase considerably.  A central jet veto, e.g.
\bq
p_{Tj} > \text{15 GeV}\quad \text{and}\quad |\eta_j| < 3.2\;,
\label{centraljetvetodef}
\eq
is very effective in suppressing the \ttb background
to inclusive $H\to WW$ searches at the LHC
\cite{DittDrein,CMS,ATLAS_TDR_2,Trefzger_Higgs}.  The veto of 
Eq.~(\ref{centraljetvetodef}) reduces the inclusive cross section
of 6 pb to 14 fb when the NWA or DPA is applied, whereas a
calculation with complete matrix elements yields 26 fb.  Sub- and
non-resonant contributions increase the result by a factor 1.8.
As in the inclusive case, moving from NWA to DPA
or switching finite-width schemes changes the corresponding result very
little in comparison.  The distributions in Fig.~\ref{JetVetoDistributions} show
this relationship for differential cross sections.

\begin{table}
\caption{\label{LightHiggsDiscoveryOffshellTopBackgrounds}
Top background cross sections with up to two additional jets
for the $H\to WW\to e^\pm\mu^\mp \sla{p}_T$ decay search in weak
boson fusion at the LHC.  The light Higgs-optimised selection
cuts and event classification from Ref.~\cite{KPRZ} are applied.
All cross sections are given in fb.
Sub- and non-resonant amplitude contributions enhance the total
top background by a factor of 1.1.
The \ttb background without additional jets is strongly
suppressed because in this case both $b$ quarks need to be
resolved as forward jets with wide separation in pseudorapidity
and very large di-jet invariant mass.
}
\vspace*{.5cm}
\begin{tabular}{|l|c|c||c|c||c|c|}
 \cline{2-7}
\multicolumn{1}{c|}{} & \multicolumn{2}{c||}{$t\bar{t}$} & \multicolumn{2}{c||}{$t\bar{t}j$} & \multicolumn{2}{c|}{$t\bar{t}jj$} \\
 \cline{2-7}
\multicolumn{1}{c|}{} & $\sigma$ & $\sigma/\sigma_\text{NWA}$ & $\sigma$ & $\sigma/\sigma_\text{NWA}$ & $\sigma$ & $\sigma/\sigma_\text{NWA}$ \\
  \hline
NWA  & 0.020 & 1.0  & 0.94 & 1.0 & 0.24 & 1.0 \\
FWS  & 0.044 & 2.1  & 1.08 & 1.1 & 0.24 & 1.0 \\
CMS  & 0.044 & 2.1  & 1.07 & 1.1 & 0.24 & 1.0 \\
OFS  & 0.044 & 2.1  & 1.07 & 1.1 & 0.24 & 1.0 \\
\hline 
\end{tabular}
\end{table}

$H\to WW$ searches are usually tuned for intermediate Higgs masses around 170 GeV,
where the $H\to WW$ branching ratio is large.
In Ref.~\cite{KPRZ}, the search for $H\to WW$ decays in weak boson fusion
at the LHC is studied for the light Higgs ($m_H = 115$ GeV) favoured by
LEP experiments \cite{lep2higgs}.  The additional forward jets in weak boson
fusion permit powerful selection cut optimisations that make this search channel
competitive -- a $5\sigma$ discovery is possible with
35 fb$^{-1}$ -- even for relatively low Higgs masses where the $H\to WW$ branching
ratio is small.  In this search scenario, \ttb + jets production is the dominant
background and its accurate determination is essential.
The \ttb background is strongly suppressed, because for final states without
additional jets both $b$ quarks need to be resolved as forward jets with wide
separation in pseudorapidity and very large di-jet invariant mass.
In Ref.~\cite{KPRZ}, complete matrix element corrections were calculated
for the \ttb and $t\bar tj$ backgrounds using the OFS.  As shown in Section
\ref{BeyondSM}, OFS cross sections can be artificially reduced.  Our program
allows to calculate these corrections using the FWS, CMS or OFS.
The results are given in Table \ref{LightHiggsDiscoveryOffshellTopBackgrounds}
and show that the OFS is reliable in this case.  
Table \ref{LightHiggsDiscoveryOffshellTopBackgrounds} also displays first
results for $t\bar t$+2 jets production calculated with complete LO matrix elements
in the literature.
Sub- and non-resonant amplitude contributions enhance the total
top background by a factor of 1.1.


\subsection{Backgrounds to Beyond-SM Physics Searches}
\label{BeyondSM}

At the LHC, top backgrounds also play an important part in searches for
physics beyond the Standard Model.  In this section we focus on two studies
where \ttb production constitutes the dominant background: the search for
hadronic $\tau$ decay of a heavy charged Higgs 
in supersymmetric models and the radion decay $\phi\to hh\to b\bar{b}\tau^+\tau^-$,
where one $\tau$ decays leptonically and the other hadronically.

\begin{table}
\caption{\label{ChargedHiggsOffshellTopBackgrounds}
Top background for heavy charged Higgs production
$gg\to H^\pm tb$
and decays $H^\pm\to \tau\nu$ (with hadronic $\tau$ decay)
and $t\to jjb$ at the LHC.
The selection cuts of the ATLAS analysis in
Ref.~\protect\cite{chargedHiggs} are applied.
Sub- and non-resonant amplitude contributions enhance the 
top background by a factor of 1.1.
The ATLAS analysis combines \ttb and $Wtb$ results in NWA
at the cross-section level.
This procedure can lead to substantial double-counting of
sub- and non-resonant contributions \protect\cite{ttj}, evidently a 30\% effect in the case at hand.
All cross sections are given in fb.  Parton-level results are
rescaled by a factor 0.16, so that our NWA result and the PYTHIA-ATLFAST
\cite{pythia,atlfast} \ttb result given in Table 3 in
Ref.~\protect\cite{chargedHiggs} match.
}
\vspace*{.5cm}
\begin{tabular}{|l|c|c|}
 \cline{2-3}
 \multicolumn{1}{c|}{} & $\sigma$ & $\sigma/\sigma_\text{NWA}$ \\
 \hline
 NWA & 0.343 & 1.00 \\
 NWA $(t\bar{t}\;+\;Wtb)$\footnote{calculated in Ref.~\protect\cite{chargedHiggs}} & 0.485 & 1.41 \\
 FWS & 0.376 & 1.09 \\
 CMS & 0.378 & 1.10 \\
 OFS & 0.364 & 1.06 \\
 \hline 
\end{tabular}
\end{table}

The production of a charged Higgs with 
$m_{H^\pm} > m_t$ in supersymmetric models at high $\tan \beta$
was analysed in Ref.~\cite{chargedHiggs}.
Production proceeds through $gg\to H^\pm tb$ and is followed by the
decays $H^\pm\to \tau\nu$ (with hadronic
$\tau$ decay) and $t\to jjb$.
Applying the selection cuts of this ATLAS study, we calculate \ttb background
cross sections in NWA and with complete matrix elements to determine the
enhancement factor.  The results are shown in Table
\ref{ChargedHiggsOffshellTopBackgrounds}.
Sub- and non-resonant amplitude contributions enhance the 
top background by a factor of 1.1.
The ATLAS analysis takes sub-resonant contributions into account by
combining \ttb and $Wtb$ results in NWA at the cross-section level.
This procedure can lead to substantial
double-counting of sub- and non-resonant contributions \cite{ttj}.
Our enhancement factor indicates that the actual top background is 23\%
lower than the estimate in Ref.~\cite{chargedHiggs}.
In Ref.~\cite{chargedHiggsExtraDim}, the analysis was extended to 
the search for $H^-\to \tau_L^-\nu$ in models with a singlet
neutrino in large extra dimensions, and we expect a similarly reduced
top background if sub-resonant contributions are included at the
amplitude-level.

\begin{table}
\caption{\label{RadionTauTauOffshellTopBackgrounds}
Top background for the radion decay
$\phi\to hh\to b\bar{b}\tau^+\tau^-$ at the LHC,
where one $\tau$ decays leptonically and the other hadronically.
The selection cuts of the ATLAS analysis in
Ref.~\protect\cite{radiontautau} are applied.
The radion vacuum expectation value $\Lambda_\phi = 1$ TeV,
the radion-SM Higgs mixing parameter $\xi = 0$, the radion mass
$m_\phi = 300$ GeV and the lightest Higgs mass $m_h = 125$ GeV.
Our results indicate that sub- and non-resonant
amplitude contributions change the top background by not more than 3\%.
All cross sections are given in fb.  Parton-level results are
rescaled, so that our NWA result and the PYTHIA-ATLFAST \cite{pythia,atlfast}
\ttb result given in Table 5 in Ref.~\protect\cite{radiontautau} match.
}
\vspace*{.5cm}
\begin{tabular}{|l|c|c|}
 \cline{2-3}
 \multicolumn{1}{c|}{} & $\sigma$ & $\sigma/\sigma_\text{NWA}$ \\
 \hline
 NWA & 3.27 & 1.00 \\
 FWS & 3.36 & 1.03 \\
 CMS & 3.34 & 1.02 \\
 OFS & 3.17 & 0.97 \\
 \hline 
\end{tabular}
\vspace*{.3cm}
\end{table}

In Table \ref{RadionTauTauOffshellTopBackgrounds}, enhancement factors
are given for the top background to the decay
$\phi\to hh\to b\bar{b}\tau^+\tau^-$ of a Randall-Sundrum radion with mass 300 GeV.
The two $\tau$ leptons decay leptonically and hadronically, respectively.
The selection cuts of the ATLAS analysis in Ref.~\protect\cite{radiontautau}
are applied.  Specific model parameters are given in the
table caption.
The effect of sub- and non-resonant amplitude contributions is
small for this top background, in fact smaller than for inclusive \ttb
production.

\begin{table}
\caption{\label{OFSDeviations}
Cross sections for \ttb production with di-lepton decay
($b\bar{b}\: e^+ \nu_e\:\mu^- \bar{\nu}_\mu$) at the LHC, 
calculated in DPA and with complete matrix elements using
several finite-width schemes.  Only the phase space region where
$|m_{WW} - m_Z| < 2\Gamma_Z$ and $|m_{Wb} - m_t| < 2\Gamma_t$
is integrated.  In this region $Z$ boson {\em and} top quark propagators
are resonant.  The DPA is excellent, i.e.~resonant \ttb production
dominates.  The contribution from amplitudes with a resonant $Z$
propagator is negligible.  In this phase space region OFS matrix
elements are inadequate, since the dominant double-resonant top
amplitudes are artificially suppressed by the overall factor
$|(p_Z^2 - m_Z^2)/(p_Z^2 - m_Z^2 + im_Z\Gamma_Z)| \ll 1$.  The OFS 
result is 30\% smaller than the CMS and FWS results.
All results are given in fb.  
}
\vspace*{.5cm}
\begin{tabular}{|l|c|c|}
 \cline{2-3}
 \multicolumn{1}{c|}{} & $\sigma$ & $\sigma/\sigma_\text{DPA}$ \\
 \hline
 DPA & 0.0168 & 1.00 \\
 FWS & 0.0170 & 1.01 \\
 CMS & 0.0170 & 1.01 \\
 OFS & 0.0118 & 0.70 \\
 \hline 
\end{tabular}
\end{table}

The OFS cross sections in Tables \ref{ChargedHiggsOffshellTopBackgrounds} and 
\ref{RadionTauTauOffshellTopBackgrounds} are several percent lower than the
corresponding FWS and CMS cross sections, which agree within the integration
error of 1\%.
To understand why the OFS may not be suitable for all \ttb background
calculations, we integrate
the small phase space region where both top quarks and the intermediate $Z$ boson are 
close to resonance.  More specifically, we require $|m_{Wb} - m_t| < 2\Gamma_t$
and $|m_{WW} - m_Z| < 2\Gamma_Z$.
The results are shown in Table \ref{OFSDeviations}.  The DPA is excellent
and contributions from amplitudes with resonant
$Z$ propagator are negligible.  In the OFS, the dominant double-resonant top
amplitudes are artificially suppressed by the overall factor
$|(p_Z^2 - m_Z^2)/(p_Z^2 - m_Z^2 + im_Z\Gamma_Z)| \ll 1$.  The 
resulting OFS cross section is consequently much smaller than CMS or FWS cross
sections.  This example illustrates that cross sections for
multi-resonant processes cannot be calculated reliably with the
OFS if sizable contributions arise from phase space regions where
several amplitudes with different resonance structure compete.
Artificially reduced cross sections can even occur for single-resonant
processes, given that non-resonant contributions are sizable in phase space
regions close to resonance.  The authors of Ref.~\cite{Baur:1995aa}, for
example, compared OFS and fermion-loop scheme cross sections for radiative $W$
production and found that OFS results are 30\% lower close to threshold.
Despite some effort\footnote{We considered, for example, the
  selections $m_{Wb} > 500$ GeV, $|\eta_b| > 3.5$ and $\Delta
  R_{WW} < 1.0$ for collider energies up to 100 PeV.}, we were unable to
find a phase space region, where CMS and FWS cross sections for \ttb
production showed significant discrepancies.  We therefore conjecture that
calculations employing the gauge-variant fixed-width scheme may be
used to obtain reliable predictions for the processes considered
here.  We note that the reliability of the fixed-width scheme
has recently also been established for $e^+e^-\to 6$ fermion processes \cite{LUSIFER}.

Our LO calculations do not include log-enhanced higher-order contributions
from collinear $g\to b\bar b$ configurations for initial state gluons.
An improved treatment would include $gb$ scattering
matrix elements convoluted with the $b$ quark PDF.  Then, a subtraction
of the gluon splitting term would also be required to avoid
double-counting \cite{gbContribution}.  However, the
additional net contribution to inclusive \ttb production
is less than 2\% \cite{Boos_SingleTop_Followup}
and can safely be neglected in our analysis in Section \ref{inclusive}.
In the weak boson fusion Higgs search discussed in 
Section \ref{SMSearch}, one or both $b$ quarks have no finite
transverse momentum threshold, but collinear contributions are
small within typical cuts \cite{ttj}.
For central jet veto suppressed top backgrounds, on the other
hand, one would expect more pronounced collinear enhancement.
The selection cuts for the \ttb background studies in Section
\ref{BeyondSM} require that the $b$ quarks are resolved with a
transverse momentum of at least 15 GeV.  The collinear region is
thus avoided.


\section{\label{summary}Summary and Outlook}

We presented cross sections for top quark pair production at hadron colliders
with up to two additional jets resulting in 6-, 7- and 8-particle final states
calculated with complete tree-level matrix elements.  Our program includes
di-lepton, single-lepton and all-jet decay modes, and implements three
practical finite-width schemes, i.e.~the fixed-width, complex-mass and
overall-factor schemes, as well as the narrow-width and double-pole
approximation for comparison.
While our LO calculations are subject to substantial scale uncertainties, the obtained cross section ratios are expected to be robust.
For inclusive production, advancing from NWA to DPA by replacing on-shell with
off-shell intermediate top quarks, changes cross sections by 1\% or
less.  The inclusion of sub- and non-resonant amplitudes increases NWA or DPA
cross sections by 5--8\% at the LHC and VLHC, but has little effect on Tevatron
cross sections.
Top backgrounds to new particle searches are often suppressed by
optimised selection cuts that can enhance the importance of sub- and
non-resonant contributions considerably.  We updated on-shell $t\bar t/Wtb$ background
estimates for the $H\to WW$ decay in weak boson fusion, the hadronic
$\tau$ decay of a heavy $H^\pm$ and the $\phi\to hh\to\tau\tau b\bar{b}$ radion
decay at the LHC, and found corrections from 3\% to 30\%.
All calculated FWS and CMS cross sections agree within errors.
Gauge-violating effects of the FWS appear to be generally negligible for the
processes considered here.  Our calculations show further that the OFS
may yield underestimated cross sections and should be applied with caution
in studies with suppressed top backgrounds.

Because of the large scale uncertainties of LO cross sections,
precise absolute predictions for top pair production at hadron colliders
cannot be achieved with tree-level calculations.  
The extension of LO to NLO calculations in the framework of the narrow-width
and double-pole approximations was first explored in the context of weak gauge
boson production \cite{WWinNLOwithDPAforLinColl} and has recently also been
carried out for top pair production at hadron colliders
\cite{ttinNLOwithDPAforHadrColl}.  The results in Section \ref{inclusive}
imply that sub-resonant contributions need to be included in NLO calculations for 
inclusive \ttb production at $pp$ supercolliders to 
achieve a theoretical error of ${\cal O}(5\%)$.
A common method to improve LO predictions for suppressed top backgrounds
is to apply a reaction-specific $K$-factor, i.e.~to rescale all
LO results by multiplying
with $K = \sigma_{incl,\,\text{NLO}}/\sigma_{incl,\,\text{LO}}$.  When
sub-resonant and non-resonant phase space regions contribute substantially to
cross sections, the merit of such procedures has to be tested by 
comparing with fully differential NLO calculations that
cover resonant and non-resonant phase space regions.
The starting point for a complete NLO
calculation of top pair production, i.e. a calculation that is not specialised
to the double-resonant phase space region, would be the evaluation of the
NLO corrections of the complete matrix element for the $b\bar{b}W^+W^-$ final
state.  The calculation of the real emission corrections is straightforward,
since the $W$ bosons are on-shell.
However, the evaluation of the virtual corrections of this
$2\to 4$ process involves 1-loop hexagon amplitudes, whose computation
is still very challenging \cite{Hexagon}.  While the $t\bar t +\,\text{jets}$
program described in this paper allows to calculate the real emission
component of a complete calculation of $pp\to$ 6 fermions at NLO in QCD,
the evaluation of the virtual corrections for such $2\to 6$ processes is well
beyond present capabilities.  NLO predictions for many-particle
processes with multiple scales can be further improved by
resumming higher-order contributions with large logarithms, such as $\alpha_s
\log(m_t^2/p^2_{Tj})$ in the case at hand.

In addition to precise and accurate calculations for hard scattering
subprocesses, a reliable comparison of theoretical predictions and experimental
data also requires the proper inclusion of parton showering, hadronisation and
detector effects.  To standardise the co-operation of parton-level Monte Carlo
programs (with full matrix elements) and showering and hadronisation event
generators -- which in turn produce input for detector simulations --
a generic interface has been specified recently in
Ref.~\cite{LesHouchesEventGeneratorInterface}.  In the near future, we plan to
implement this interface and to make our complete LO top pair production program
available to interested experimental physicists.

\begin{acknowledgments}
We thank T.~Stelzer for access to a recent version of MADGRAPH and T.~Trefzger
for helpful information.
\end{acknowledgments}

\end{document}